\documentclass[a4paper,11pt]{article}
\usepackage{pos}  

\newcommand{\ubl}{U(1)_{B_3-L_2}}
\newcommand{\bsmm}{${b \rightarrow s \mu^+ \mu^-}$ }
\newcommand{\hc}{^{\dagger}} 
\let\originalleft\left
\let\originalright\right
\renewcommand{\left}{\mathopen{}\mathclose\bgroup\originalleft}
\renewcommand{\right}{\aftergroup\egroup\originalright}

\usepackage{mathtools}
\usepackage{amscd}
\usepackage{amsmath}
\usepackage{slashed}
\usepackage{bm}

\renewcommand{\logo}{\relax}

\title{Searching for the flavon at current and future colliders}

\author*[a]{Eetu Loisa}

\affiliation[a]{DAMTP, University of Cambridge,\\
  Wilberforce Road, Cambridge, CB3 0WA, United Kingdom}

\emailAdd{eal47@cam.ac.uk}

\abstract{
The $ B_3 - L_2$ $ Z' $ model may explain certain features of the fermion mass spectrum as well as the \bsmm anomalies.
The $ Z' $ acquires its mass via a TeV-scale scalar field, the flavon, whose vacuum expectation value spontaneously breaks the family non-universal gauged $ \ubl $ symmetry.
We review the key features of the model, with an emphasis on its scalar potential and the flavon field, and use experimental data and perturbativity arguments to place bounds upon the Higgs--flavon mixing angle.
Finally, we discuss \textit{flavonstrahlung} as a means to discover the flavon experimentally and compute flavonstrahlung cross-sections at current and future colliders.

}

\FullConference{21st Conference on Flavor Physics and CP Violation (FPCP 2023)\\
 29 May - 2 June 2023\\
 Lyon, France\\}

\begin{document}
\maketitle

\section{Introduction}
Several observables involving rare $ B $ meson decays with muonic final states remain in disagreement with Standard Model (SM) predictions. 
For instance, discrepant branching ratios of $ B \rightarrow K \mu \mu$, $ B \rightarrow K^* \mu \mu $  and $ B_s \rightarrow \phi \mu \mu $ \cite{LHCb:2014cxe,LHCb:2016ykl,LHCb:2021zwz} as well as angular observables of $ B \rightarrow K^* \mu \mu $ \cite{LHCb:2020lmf} are hinting that there may be New Physics (NP) at play.  
These measurements can be accounted for by the $ B_3 - L_2 $ model \cite{Alonso:2017uky,Bonilla:2017lsq,Allanach:2020kss}, which introduces a family non-universal $ \ubl $ abelian gauge symmetry, mediated by a $ Z' $ gauge boson able to contribute to \bsmm transitions.
Whilst the recent LHCb update on the lepton flavour universality ratios $ R_K $ and $ R_{K^*}$ \cite{LHCb:2022qnv,LHCb:2022zom} guides us towards NP scenarios coupling to both muons and electrons \cite{Alguero:2023jeh,Allanach:2023uxz}, the one-parameter fits based solely on the Wilson coefficient $ \mathcal{C}_{9\mu} $ are still able to improve considerably upon the SM $ \chi^2 $ \cite{Allanach:2022iod}.
Owing to its family non-universal nature, the $ B_3 - L_2$ model is also able to explain certain features of the CKM matrix, setting up the foundations for further model-building to address the fermion mass problem. 

Since the $  Z'$ is assumed to be massive, we are compelled to introduce a scalar flavon field that breaks the $ \ubl $ symmetry by developing a vacuum expectation value (VEV).
In what follows, we will review the construction of the $ B_3 - L_2 $ and study the phenomenology of the flavon field with an eye on the Higgs--flavon mixing. 
Finally, we will study the production of the flavon at hadron and muon colliders.
This write-up draws heavily from ref.~\cite{Allanach:2022blr}, which the interested reader can refer to for a more detailed account of our findings.

\section{\boldmath{$B_3-L_2$} model}
To construct the $ B_3 - L_2 $  model, one begins by extending the $SU(3) \times SU(2)_L \times U(1)_Y$ gauge group of the SM by an abelian $ \ubl $ factor in a direct product.
The SM fermions carry charges proportional to third family baryon number ($ B_3 $) minus second family lepton number ($ L_2 $) under the new symmetry. 
The exact charge assignments are shown in table~\ref{tab:charges}.
Gauge anomaly cancellation is ensured by assuming the existence of three right-handed neutrinos.
Because the $ \ubl $ symmetry is gauged, it implies the existence of an electrically neutral $ Z' $ boson which is able to contribute to flavour-changing neutral current (FCNC) process, including  \bsmm  transitions.
We also introduce a complex scalar field called the flavon ($ \theta$), which is a SM singlet but carries charge $ q_\theta $ under $ \ubl $. 
(We assume $ q_\theta = 1 $ in this work.) 
The flavon field develops a non-zero VEV near the TeV scale, $ \langle \theta \rangle = v_\theta \neq 0 $. 
This breaks $ \ubl $ and yields the $ Z' $ mass $ M_{Z'} = q_\theta g_{Z'} v_\theta $, where $ g_{Z'}  $ is the coupling constant of the new gauge symmetry. 

\begin{table}[htpb]
	\centering
	\begin{tabular}{ccccccccc}
		\hline \hline
		$ Q'_{iL} $ & $ u'_{iR} $ & $ d'_{iR} $ & $ L_1' $& $ L_2' $ & $ L_3' $ & $ e'_{1 R} $ & $ e'_{2R} $ & $ e'_{3R} $ \\
		0 & 0 & 0 & 0 & -3 & 0 & 0 & -3 & 0  \\
		\hline
		$\nu'_{1R} $ & $\nu'_{2R} $ & $\nu'_{3R} $ & $ Q'_{3L} $ & $ u'_{3R} $ & $ d'_{3R} $ & $ H $ & $ \theta $ \\
	0 & -3 & 0 & 1 & 1 & 1 & 0 & $ q_{\theta} $ \\
	\hline \hline
	\end{tabular}
	\caption{The $\ubl$ charge assignments. A prime stands for a weak
          eigenstate Weyl fermion and the family index $ i $ takes values 1
          and 2. The flavon charge $ q_{\theta} $ is a non-zero rational
  number set equal to unity in this work. \label{tab:charges}}
\end{table}

The fermionic couplings of the $ Z^\prime $ are expressed as
\begin{equation} \label{eq:zprime_fermion_lagr_weak}
	\mathcal{L}_{Z^\prime \psi} =
	- g_{Z^\prime} \left( 
		\overline{Q'_{3L}} \slashed{Z}^\prime Q'_{3L} 
	+ \overline{u'_{3R}} \slashed{Z}^\prime u'_{3R} 
+  \overline{d'_{3R}} \slashed{Z}^\prime d'_{3R}  
- 3 \overline{L'_{2L}} \slashed{Z}^\prime L_{2L}' 
- 3 \overline{e'_{2R}} \slashed{Z}^\prime e'_{2R}  
- 3 \overline{\nu'_{2R}} \slashed{Z}^\prime \nu'_{2R} \right) 
\end{equation}
where the primed fermion fields are in the weak eigenbasis.
In order to obtain phenomenologically useful results we must transform to the (unprimed) mass eigenbasis. 
Expressing the three-component column vectors in family space as boldface letters, the transformation between the two bases is written as
\begin{equation}
	\mathbf{P}'_I = V_I \mathbf{P}_I
\end{equation}
for $ I \in \left\{ u_L, d_L, e_L,\nu_L, u_R, d_R,e_R, \nu_R \right\}  $. 
Owing to the family universality of its gauge interactions, the SM is not sensitive to the form of the individual mixing matrices $ V_I $. 
The model builder is allowed to tune the $ V_I $ to their liking as long as the matrices are unitary and reproduce the measured CKM and PMNS matrix elements, given by $ V_{\text{CKM}} = V_{u_L}\hc V_{d_L} $ and $ U_{\text{PMNS}} = V_{\nu_L}\hc V_{e_L} $.
We pick a simple ansatz which is able to mediate \bsmm transitions by turning on the $ \mathcal{C}_{9\mu} $ Wilson coefficient and not ruled out by tight experimental bounds on FCNC processes: 
\begin{equation}
V_{d_L} = \begin{pmatrix} 
	1 & 0 & 0 \\
	0 & \cos \theta_{sb} &  - \sin \theta_{sb}\\
	0 & \sin \theta_{sb}& \cos \theta_{sb} \\
	\end{pmatrix},
\end{equation}
$ V_{d_R} = 1, V_{e_R} = 1, V_{e_L}=1 $ and $ V_{u_R} = 1$, where $1$
denotes the 3 by 3 identity matrix. These imply $V_{u_{L}} = V_{d_L} V\hc $ and $ V_{\nu_L} = U\hc $.
Substituting the ansatz into Eq.~\ref{eq:zprime_fermion_lagr_weak}, one finds the terms
\begin{equation}
	\mathcal{L}_{Z' \psi} \supset - g_{Z^\prime} \left[\left(  \frac{1}{2} \sin 2 \theta_{sb} \overline{s} \slashed{Z}^\prime P_L b + \text{H.c.}  \right) - 3 \overline{\mu} \slashed{Z}^\prime \mu \right] 
\end{equation}
which, upon integrating out the $ Z' $, will contribute to the $ \mathcal{O}_{9 \mu} $ operator in the WET Hamiltonian 
\begin{equation}
	\mathcal{H}_{\text{WET}} \supset \mathcal{C}_{9\mu} \mathcal{N} \left( \overline{s} \gamma_\nu P_L b  \right) \left( \overline{\mu} \gamma^\nu \mu \right). 
\end{equation}
with $ \mathcal{C}_{9 \mu} \sim g_{Z'}^2 \sin 2 \theta_{sb} / M_{Z'}^2 $ and $ \mathcal{N} $ a model-independent constant.
For each choice of $ g_{Z'} $ and $ M_{Z'} $, we tune $ \theta_{sb} $ such that this expression matches the value of $ \mathcal{C}_{9 \mu} $ obtained from fits to \bsmm data.\footnote{We have used the best-fit value obtained in \cite{Altmannshofer:2021qrr} before the LHCb updates \cite{LHCb:2022qnv, LHCb:2022zom} of the lepton flavour universality ratios $ R_{K} $ and $ R_{K^*} $, $ \mathcal{C}_{9\mu} = -0.73 \pm 0.15 $.
The results presented here are very weakly dependent on the value of $ \theta_{sb} $ and using an up-to-date determination of $ \mathcal{C}_{9\mu} $ would not have an observable impact on them.}
This procedure eliminates $ \theta_{sb} $ as an independent parameter.

The $ \ubl $ symmetry places restrictions on the forms of the Yukawa matrices of the model, causing the CKM matrix to take the form 
\begin{equation}
 	V_{\text{CKM}}^{\text{model}} \sim 
		\begin{pmatrix} 
			\times & \times & 0 \\
			\times & \times & 0 \\
			0 & 0 & \times \\
		\end{pmatrix}
\end{equation}
at the renormalisable level.
Here $ \times $ stands for an arbitrary order one element. 
This broadly agrees with the experimentally determined CKM matrix \cite{Workman:2022ynf},
\begin{equation}
	V_{\text{CKM}}^{\text{exp.}} \approx 
		\begin{pmatrix} 
			1 & 0.2 & 0.004 \\
			0.2 & 1 & 0.04 \\
			0.009 & 0.04 & 1 \\
		\end{pmatrix}, 
\end{equation}
enabling the $ \ubl $ model to act as a bottom-up starting point in explaining the structure of the CKM matrix.

\subsection{Spontaneous symmetry breaking}\label{sec:flavon_potential}
The flavon field $ \theta $ modifies the scalar potential of the theory, which reads
\begin{equation} \label{eq:scalar_potential}
	V(H, \theta) = -\mu_{H}^2 H^{\dagger} H + \lambda_H (H \hc H )^2 - \mu_{\theta}^2 \theta^\ast \theta + \lambda_{\theta} (\theta^\ast \theta )^2 + 	\lambda_{\theta H} \theta^\ast \theta H^{\dagger} H.
\end{equation}
The last term is of particular significance as it allows the SM Higgs to interact with the flavon. 
Scalar potentials of this form have been studied in more detail in e.g.\ \cite{Barger:2008jx,Pruna:2013bma,Falkowski:2015iwa}.
Working in the unitary gauge and expanding both fields about their VEVs:
\begin{equation} 
	H = 
	\begin{pmatrix}
		0 \\
		\frac{v_H + h'}{\sqrt{2}}  \\
	\end{pmatrix}, 
	\qquad \theta = \frac{v_{\theta}  + \vartheta'}{\sqrt{2}},
\end{equation}
one obtains terms bilinear in the two scalars:
\begin{equation}
	V(H,\theta) \supset - \lambda_{\theta H} v_{\theta} v_H h' \vartheta',
\end{equation}
meaning that the scalar field mass matrix is non-diagonal. 
We perform a field rotation parameterised by an angle $ \phi $ to go from the (primed) non-diagonal field basis to the (unprimed) mass basis:
\begin{equation} \label{eq:rotation}
    \begin{pmatrix}
    h \\
    \vartheta \\
    \end{pmatrix}
    = 
    \begin{pmatrix}
    \cos \phi & -\sin \phi \\
    \sin \phi & \cos \phi \\
    \end{pmatrix}
    \begin{pmatrix}
    h' \\
    \vartheta' \\
\end{pmatrix}.
\end{equation}
We call $ \phi $ the Higgs--flavon mixing angle.

\section{Constraints on Higgs--flavon mixing}
\begin{figure}[htpb]
	\centering
	\includegraphics[width=0.8\textwidth]{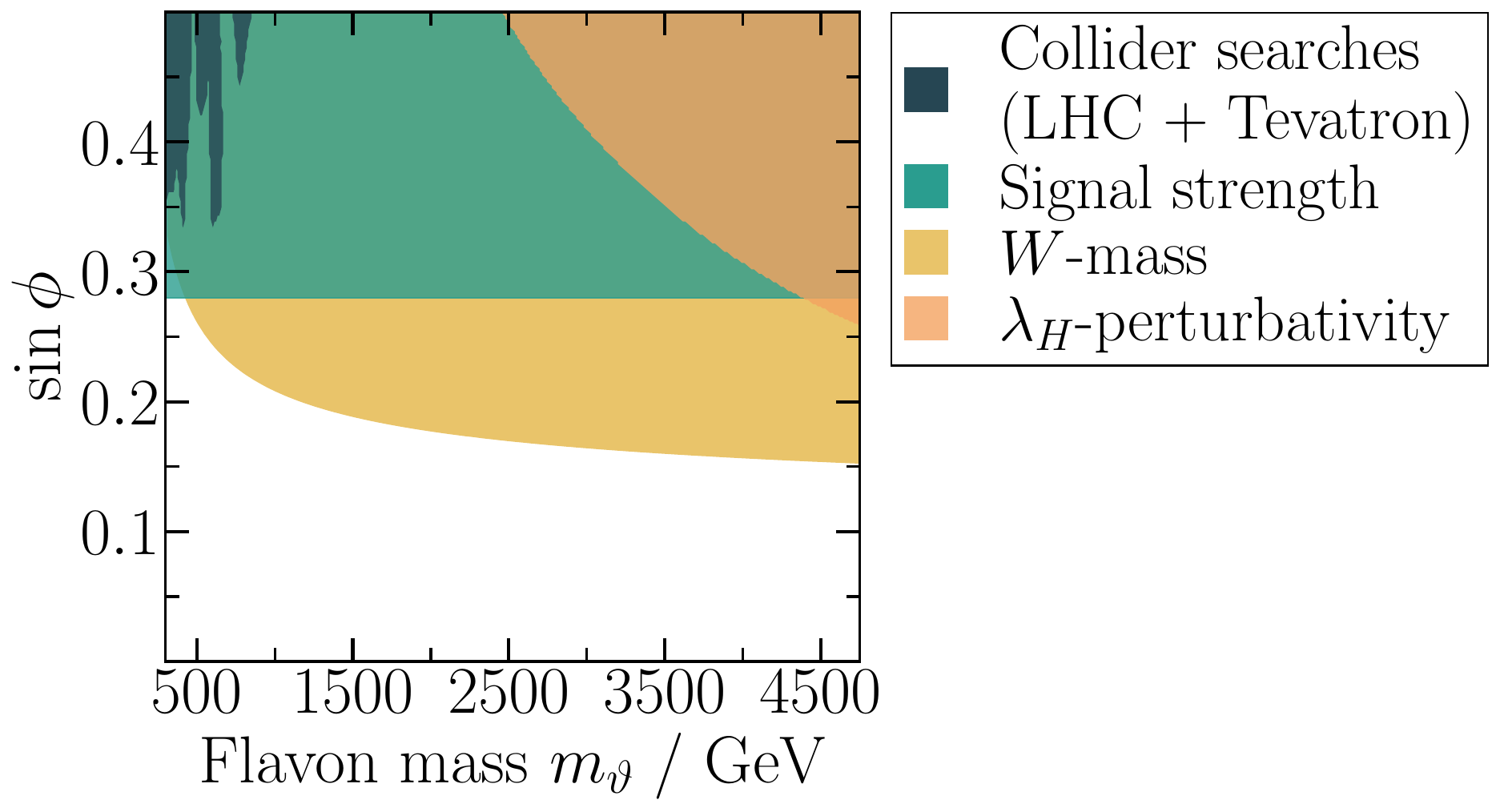}
	\caption{Various bounds on the Higgs--flavon mixing angle coming from experimental measurements and theoretical constraints. 
	The coloured regions correspond to 95\% CL search limits. 
	The dark green region comes from direct collider searches using scalar search data from the LHC and the Tevatron, whereas the light green constraint is derived from the ATLAS Higgs signal strength measurements.
	The requirement that the quartic couplings of the scalar potential be perturbative gives rise to the orange limit.
The yellow band, which is also the strictest constraint, results from insisting on agreement between the measured value of the $ W $ boson mass and the theory prediction.  
	\label{fig:mixing_constraint} }
\end{figure}

We review here the main experimental and theoretical constraints on the
Higgs--flavon mixing angle $ \phi $.
See e.g.\ refs.~\cite{Robens:2015gla,Robens:2022cun, Falkowski:2015iwa} for more detailed discussions in the context of the real singlet extension of the SM\@. 
Since most of these constraints are not affected by the $ Z^\prime $ or the flavon field being complex, we obtain constraints that largely align with the literature on the SM singlet extension. 
Here, we impose four constraints on the $ B_3 -L_2 $ model, corresponding to the four coloured regions in figure~\ref{fig:mixing_constraint}:  
\begin{enumerate}
	\item Exclusion limits from direct Higgs/scalar searches at hadron colliders, obtained using the public code \textsc{HiggsBounds} \cite{Bechtle:2020pkv}. 
	This corresponds to the dark green region in figure~\ref{fig:mixing_constraint}.
	\item Higgs signal strength measurements. The limit is obtained using the ATLAS Run 2 combination of the global signal strength \cite{ATLAS:2020qdt}. 
	This limit is shown in light green in the figure.
\item Perturbativity of the quartic couplings. Here, we require that the three quartic couplings in Eq.~\ref{eq:scalar_potential} satisfy $ \left| \lambda_{H}, \lambda_{\theta}, \lambda_{\theta H} \right| < 4 \pi $ to ensure that the model remains perturbative. 
	This requirement gives the orange region in the figure.
	\item Measurements of the $ W $ boson mass. Taking the $ Z $ boson mass $M_Z$, the Fermi constant $G_F$ and the fine structure constant $\alpha_{\text{EM}}$ as experimental inputs, the $ W $-boson mass is predicted to be 
\begin{equation}
	M_W^2 = 
	\frac{1}{2} M_Z^2 
	\left[ 1+ \sqrt{1 - \frac{4 \pi \alpha}{\sqrt{2} G_F M_Z^2} \left[1 + \Delta r (M_W^2) \right]  } \, \right].
\
\end{equation}
	where the $ \Delta r $ parameter captures SM loop effects as well as the BSM contributions coming from the flavon and the $ Z' $.
	We require that the model prediction of $ M_W $ agrees with the experimental world average (excluding the 2022 CDF measurement \cite{CDF:2022hxs}) reported by the Particle Data Group \cite{Workman:2022ynf}, which gives the yellow region in the figure.
\end{enumerate}

The figure shows that, for flavons with $ \mathcal{O}(\text{TeV}) $ masses, mixings of magnitude $ |\sin \phi| \lesssim 0.15 $ are allowed. 
The strictest constraint currently comes from the $ W $ boson mass measurements.
We note that since the $ B_3 - L_2 $ model can only ever make the $ W $ boson lighter, the model is unable to account for the 2022 CDF measurement.

\section{Flavonstrahlung at current and future colliders}
\begin{figure}[htbp]
	\centering
	\includegraphics[width=0.38\textwidth]{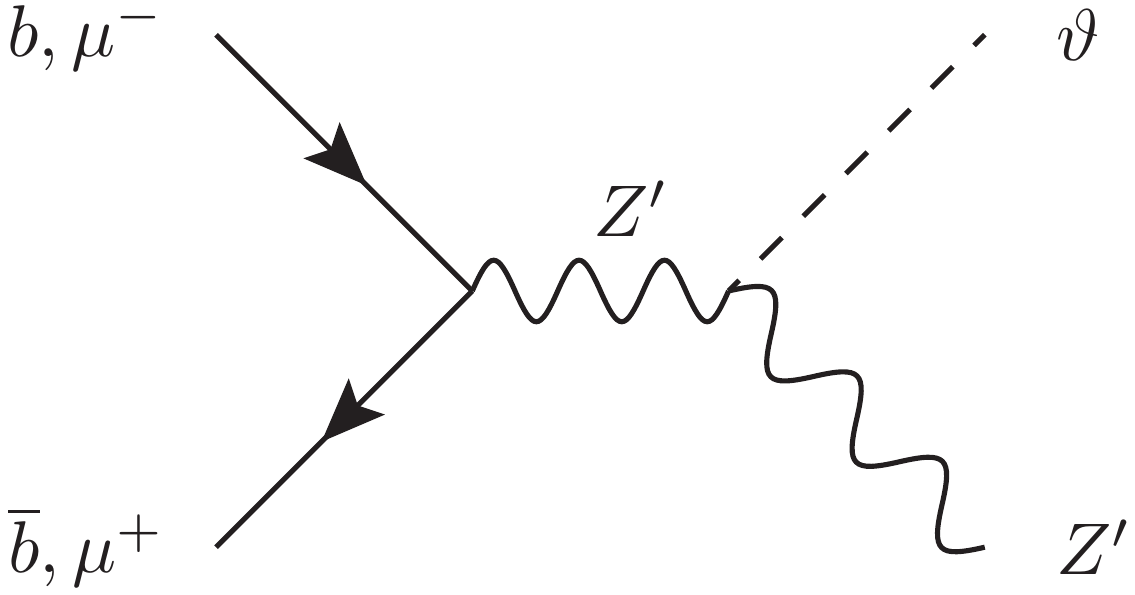}
	\caption{Flavonstrahlung at a hadron collider or a muon collider.\label{fig:flavonstrahlung}}
\end{figure}

We will now discuss the feasibility of producing the flavon at a particle collider.
To this end, we turn our attention to a process called \textit{flavonstrahlung} shown in figure~\ref{fig:flavonstrahlung}, which proceeds from a $ b \overline{b} $ or $ \mu^+ \mu^- $ initial state and leads to a final state $ Z' \vartheta $ pair (with further decays into SM particles).
The flavonstrahlung process is special compared to the conventional Higgs-like production modes of the flavon because it combines the $ Z' $ and the flavon in a single process in a way that, if observed, would confirm the role of the flavon as the source of the $ \ubl $ symmetry breaking. 
Another advantage is that, unlike the conventional Higgs-like production modes, the flavonstrahlung cross-section is proportional to $ \cos^2 \phi $ and so does not vanish in the limit of zero Higgs--flavon mixing ($ \phi \rightarrow 0 $).

\begin{figure}[htpb]
\begin{center}
\unitlength=15cm
\begin{picture}(1,0.45)(0,0)
    \put(-0.03,0.0){\includegraphics[width=0.48\textwidth]{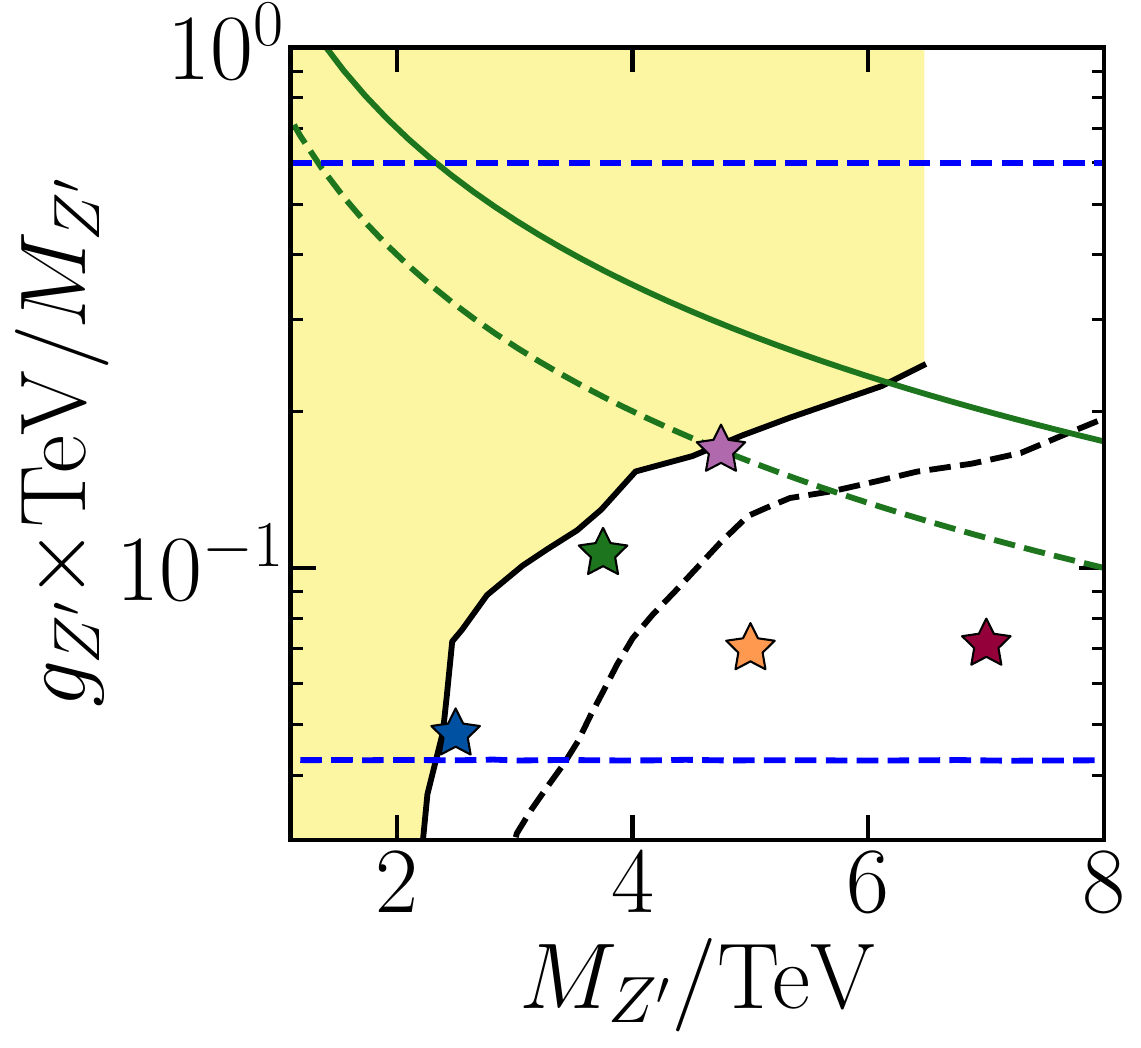}}
    \put(0.45,-0.1){\includegraphics[width=0.55\textwidth]{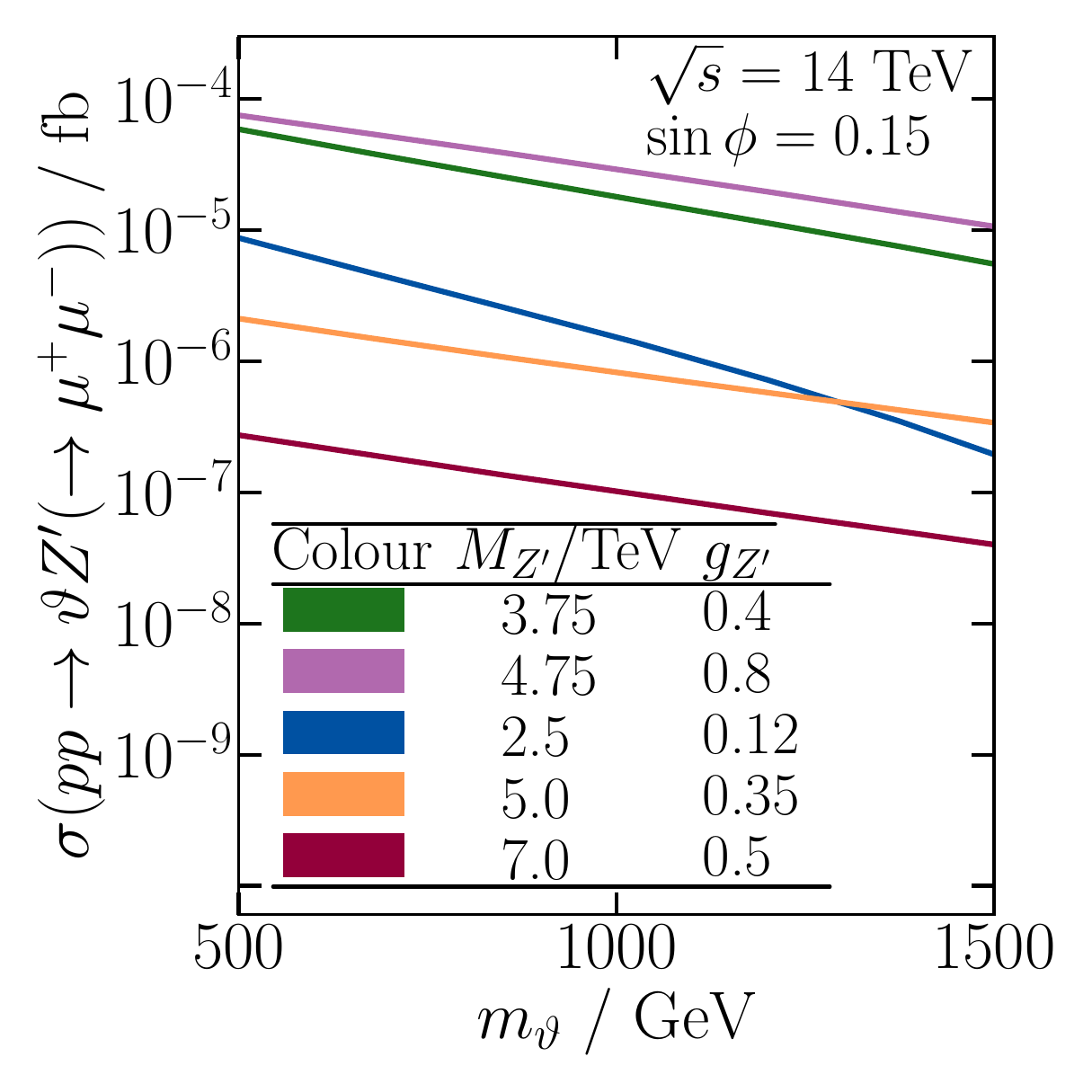}}
\end{picture}
\end{center}
\vspace*{0.8cm}
\caption{The left-hand panel, based on figure 9 of ref.~\cite{Azatov:2022itm}, shows
    the $ g_{Z^\prime} - M_{Z^\prime} $ plane of the parameter space. Everything above the
    solid black line is excluded at the 95\% CL by the LHC whereas the dashed black line
indicates the projected 95\% CL sensitivity of the HL-LHC\@. 
The dashed and solid green lines indicate the $ \Gamma / M_{Z'} = 1 / 3 $ and $  \Gamma / M_{Z'} = 1 $ bounds, above which perturbative computations become inaccurate. 
The blue dashed lines are bounds arising from the neutrino trident cross-sections and $ B_s - \overline{B_s} $ mixing; the region of parameter space between the two lines is currently allowed. Coloured stars have been superposed on the figure, with each star labelling a benchmark point in the parameter plane. The right-hand panel shows tree-level flavonstrahlung cross-sections for 14~TeV $ pp $ collisions with the flavon charge $ q_{\theta}$ set to unity. Each coloured line corresponds to a parameter space point labelled by a star of the same colour. \label{fig:14TeV_combined}}
\end{figure}

We use the event generator \textsc{\mbox{MadGraph5\_aMC@NLO}} v.3.4.1  \cite{Alwall:2014hca} to calculate leading-order flavonstrahlung cross-sections  for $ pp $ and $ \mu^+ \mu^- $ collisions $ \sigma\left( p p / \mu^+ \mu^- \rightarrow \vartheta Z^\prime  \rightarrow \vartheta \mu^+ \mu^-   \right)  $.
The final $ Z' $ is assumed to decay into a di-muon pair, a choice motivated by the large $ Z' \rightarrow \mu^+ \mu^- $ branching ratio and the clean experimental signature of the di-muon pair.
To study the process systematically, we select currently allowed combinations of $
\{M_{Z^\prime}, g_{Z^\prime}  \} $ and compute the flavonstrahlung
cross-section as a function of the flavon mass $ m_{\vartheta} $. 
These benchmark parameter choices are indicated by the five coloured stars in the left-hand panel of figure~\ref{fig:14TeV_combined}, which is adapted from figure~10 of ref. \cite{Azatov:2022itm}. The solid and dashed lines stand for various constraints on the parameter space as explained in detail in the caption of figure \ref{fig:14TeV_combined}. 
We have chosen a representative value of the Higgs--flavon mixing angle close to its upper limit, $ \sin \phi = 0.15 $, noting again that the cross-sections become smaller as $ \phi $ is increased.

We start by computing flavonstrahlung cross-sections at a centre-of-mass energy $ \sqrt{s} =$ 14~TeV, corresponding to the HL-LHC. 
The cross-sections for the five benchmark points are shown in the right-hand side of figure~\ref{fig:14TeV_combined}, where the colours of the lines correspond to the colours of the stars in the left-hand panel. 
Assuming HL-LHC integrated luminosity of 3000 $\text{fb}^{-1}$, the plot shows that less than $ \mathcal{O}(1) $ flavonstrahlung events are expected.
We conclude that the cross-sections are too small for discovery at the HL-LHC, at least when the flavon charge $ q_\theta $ is set to unity.

\begin{figure}[htpb]
	\centering
	\includegraphics[width=0.7\textwidth]{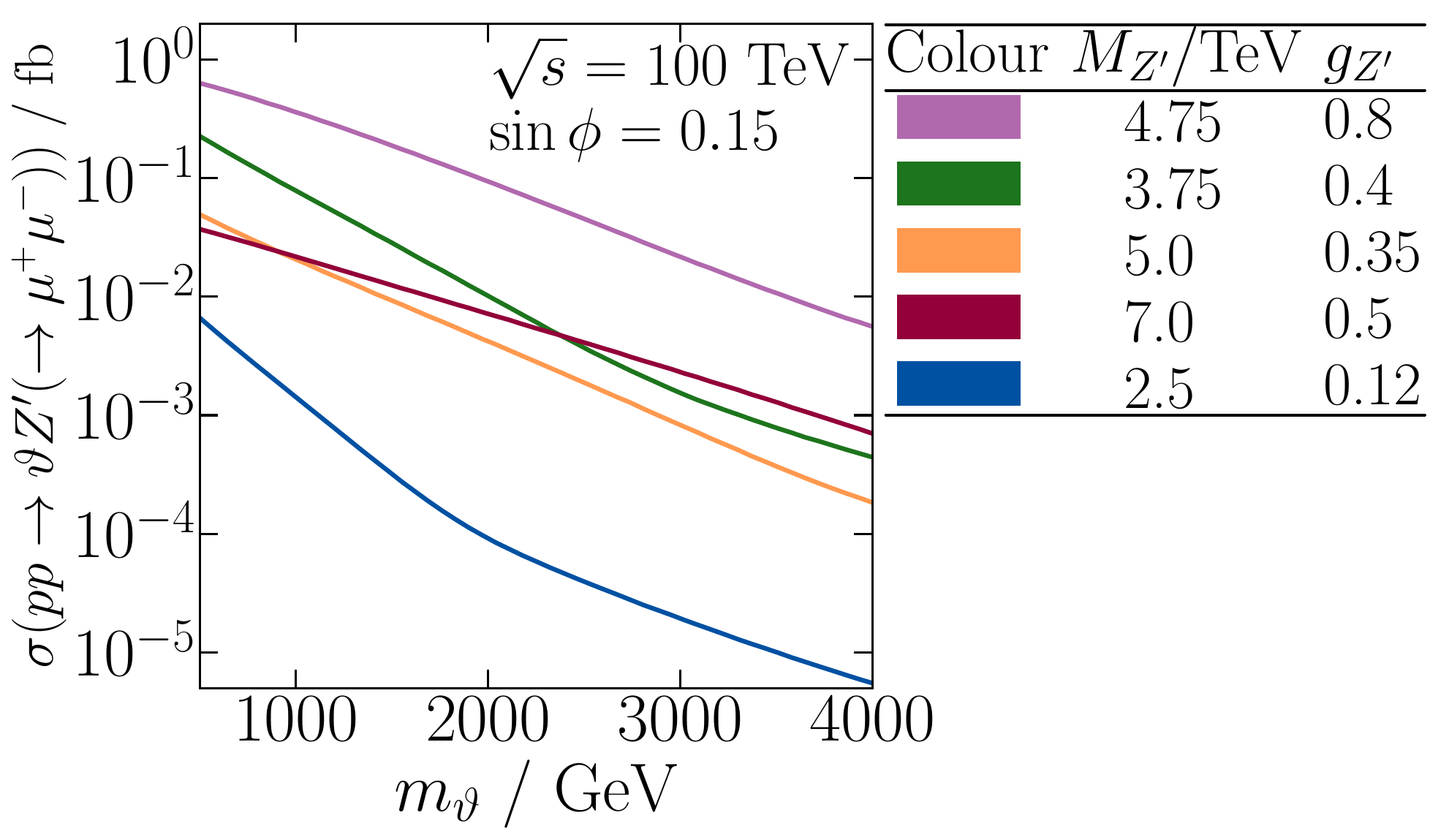}
	\caption{Tree-level flavonstrahlung cross-sections for 100~TeV $ pp $
          collisions for $q_\theta=1$. Each coloured line corresponds to
          a parameter space point labelled by a star of the same colour in
          figure~\ref{fig:14TeV_combined}. \label{fig:100TeV_combined}}
\end{figure}

Having seen that the HL-LHC lacks the centre-of-mass energy to look for flavonstrahlung, we would now like to investigate the prospects of various future colliders in observing the process.
We shall start with a 100~TeV hadron collider with an assumed  integrated luminosity of 20--30 $\text{ab}^{-1}$, representing the FCC-hh.
Figure~\ref{fig:100TeV_combined} shows the cross-sections for the five parameter space points represented by the coloured stars in figure~\ref{fig:14TeV_combined}.
The resulting cross-sections are enhanced by 3--5 orders of magnitude compared to the HL-LHC. 
To estimate the reach of the collider, we regard parameter space points at which less than 10 flavonstrahlung events are expected to be produced as undiscoverable. 
Employing this basic approach, we find that the collider can investigate the parameter space up to approximately 5 TeV flavon and $Z'$ masses, as long as $ g_{Z'} \gtrsim 0.4 $. 
For the smallest allowed values of the coupling, $ g_{Z'} \lesssim 0.4 $, the mass reach is more restricted, but the collider remains sensitive up to flavon masses of around 2~TeV.

\begin{figure}[htpb]
	\centering
	\includegraphics[width=0.7\textwidth]{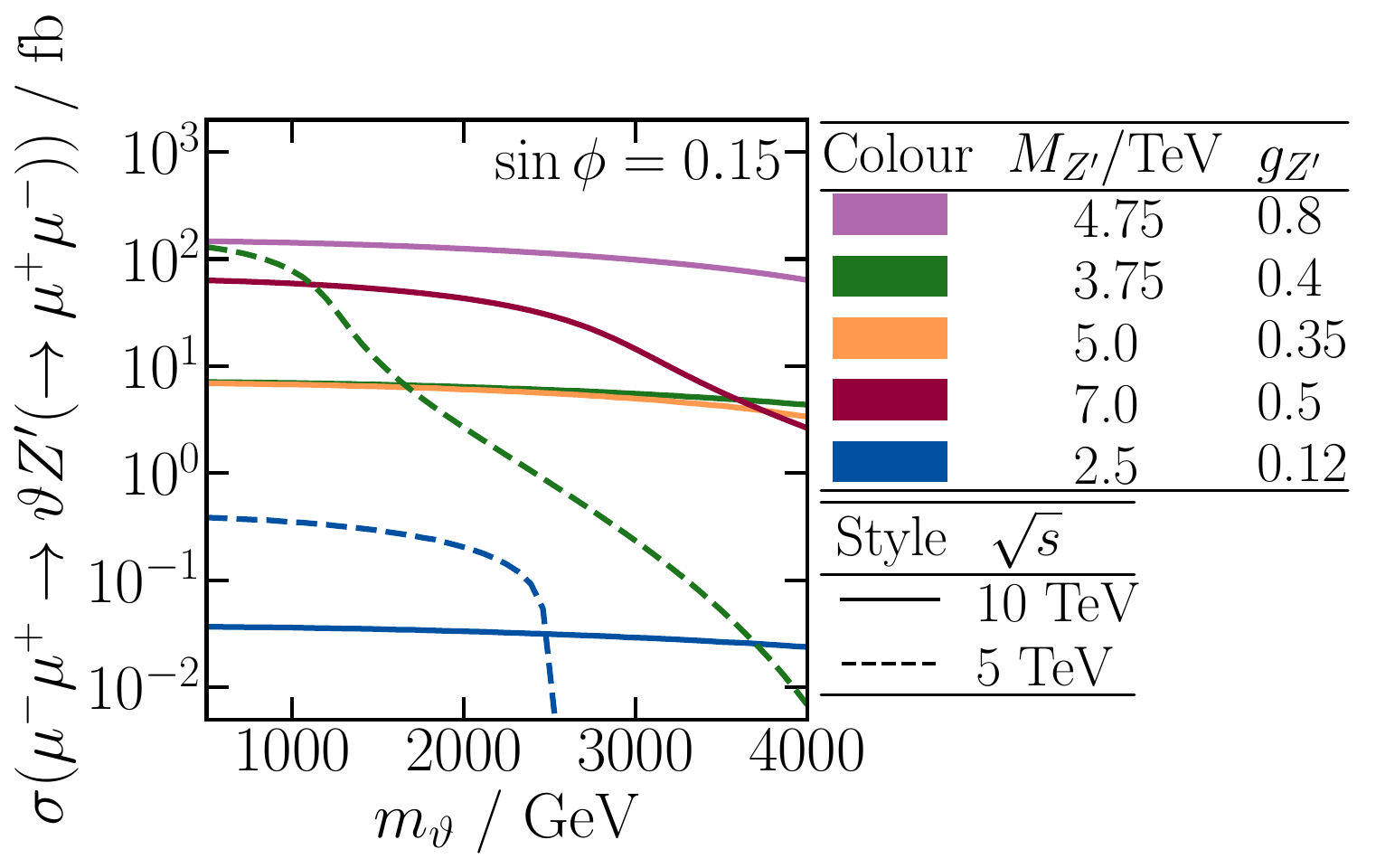}
	\caption{Tree-level flavonstrahlung cross-sections for 5~TeV (dashed lines) and 10~TeV (solid lines) $ \mu^+ \mu^- $ collisions for $ q_{\theta} = 1 $. Each coloured line corresponds to a parameter point labelled by a star of the same colour in figure~\ref{fig:14TeV_combined}. The cross-sections do not include initial state radiation effects.}
	\label{fig:muon_collider}
\end{figure}

We also simulate flavonstrahlung at 3~TeV and 10~TeV $ \mu^+ \mu^- $ colliders with integrated luminosities of 1~$\text{ab}^{-1}$ and 10 $\text{ab}^{-1}$, respectively.
The results for the five benchmark points are shown in figure~\ref{fig:muon_collider}.
The figure shows that the cross-sections at the 3 TeV collider are large enough to reach regions of the parameter space up to $ M_{Z'} \lesssim $ 5 TeV and $ m_\vartheta \lesssim 2.5 $ TeV. 
As for the 10~TeV muon collider, the figure shows that there is excellent reach up to $ M_{Z'} \lesssim 15 $ TeV and $ m_{\vartheta} \lesssim 8$ TeV. 

The flavonstrahlung cross-sections at the 10 TeV muon collider are 2--3 orders of magnitude larger than at the 100 TeV hadron collider, for a fixed parameter space point.
This outcome is to be expected since flavonstrahlung at a hadron collider requires a $ b \overline{b} $ partonic initial state, whereas the muon collider can utilise nearly the entire beam luminosity for flavonstrahlung production.

\section{Conclusions}
The $ B_3 - L_2 $ $ Z' $ model, whose salient features we have reviewed, is motivated by the \bsmm anomalies and the fermion mass puzzle.  
We have studied the flavon potential of the model and placed constraints on the size of the Higgs--flavon mixing in the model, concluding that mixing of magnitude $ | \sin \phi | < 0.15 $ is currently allowed.
The flavon may be produced through the flavonstrahlung process at hadron or muon colliders.
Whilst the cross-sections are likely too small for flavonstrahlung to be observed at the HL-LHC, we have shown that a 100~TeV hadron collider or a 10~TeV muon collider would have excellent discovery prospects.

\acknowledgments
EL would like to thank Ben Allanach for his contributions to the work presented here.
EL is supported by STFC consolidated grant ST/T000694/1. 

\bibliographystyle{JHEP-2}
\bibliography{skeleton.bib}

\end{document}